\pdfoutput=1
\documentclass[10pt,twocolumn]{article}
\usepackage[a4paper, margin=1.5cm]{geometry}
\usepackage{indentfirst}
\usepackage[utf8]{inputenc}
\usepackage[english]{babel}
\usepackage{amsmath}
\usepackage{amsfonts}
\usepackage{amssymb}
\usepackage{bm}                       
\usepackage{mathpazo}                 
\usepackage{microtype}                
\usepackage{tikz}
\usepackage{graphicx}
\usepackage{float}                    
\usepackage[export]{adjustbox}        
\usepackage[keeplastbox]{flushend}    
\usepackage[font=small,labelfont=bf]{caption}
\usepackage{booktabs}                 
\usepackage[symbol]{footmisc}         

\usepackage{hyperref}
\hypersetup{colorlinks=true, linkcolor=blue, citecolor=blue, urlcolor=blue} 

\usepackage{xcolor}
\definecolor{blue1}{RGB}{ 7,  47,  95}
\definecolor{blue2}{RGB}{18,  97, 160}
\definecolor{blue3}{RGB}{56, 149, 211}
\definecolor{red}{RGB}{210, 0, 0}

\usepackage{titlesec}                 
\titlelabel{\thetitle.\ }             
\titleformat*{\section}{\color{blue1}\scshape\bfseries\centering\large}
\titleformat*{\subsection}{\color{blue2}\normalfont\itshape\large}
\titleformat*{\subsubsection}{\color{blue3}\normalfont\itshape}
\titleformat{\paragraph}[runin]{\color{blue3}\normalfont\itshape}{}{0em}{}[~-]
\titlespacing{\paragraph}{0em}{0em}{0.3em}

\usepackage[backend=bibtex,date=year,doi=false,giveninits=true,isbn=false,maxnames=1,sorting=none,style=nature]{biblatex}
\AtEveryBibitem{
	\clearlist{language}
}
\renewcommand{\cite}[1]{\supercite{#1}}
\renewcommand{\textcite}[1]{\citeauthor{#1}\hspace*{-0.15em}\supercite{#1}}
\newlength{\spc} 
\newcommand{\citewp}[2]{
	\settowidth{\spc}{#2}
	\addtolength{\spc}{-1.8\spc}
	#2
	\hspace*{\spc}
	\supercite{#1}}

\newcommand{\snspace}[2][0.45em]{
	\hspace*{-#1}#2\hspace{0.2em}}

\begin{document}

\twocolumn[
	\begin{@twocolumnfalse}

		\begin{center}
			\textbf{\color{blue1}\large Adhesive wear and interaction of tangentially loaded micro-contacts}\\
			\vspace{1em}
			Son Pham-Ba\footnotemark\hspace*{-0.45em},\hspace{0.1em} Tobias Brink, Jean-François Molinari\\\vspace{0.5em}
			\textit{\footnotesize Institute of Civil Engineering, Institute of Materials Science and Engineering,\\\vspace{-0.2em}
			École polytechnique fédérale de Lausanne (EPFL), CH 1015 Lausanne, Switzerland}
		\end{center}


		\begin{center}
			\parbox{14cm}{\small
				\setlength\parindent{1em}Current engineering wear models are often based on empirical parameters rather than built upon physical considerations. Here, we look for a physical description of adhesive wear at the microscale, at which the interaction between two surfaces comes down to the contact of asperities. Recent theoretical work has shown that there is a critical micro-contact size above which it becomes energetically favorable to form a wear particle. We extend this model by taking into consideration the elastic interaction of multiple closely-spaced micro-contacts in 2D, with different sizes and separation distances. Fundamental contact mechanics solutions are used to evaluate the elastic energy stored by shearing the micro-contacts, and the stored energy is compared to the energy needed to detach a single joined debris particle or multiple debris particles under the micro-contacts. Molecular dynamics simulations are used to test the predictions of the outcome for various sets of parameters. Our model provides simple criteria to evaluate the energetic feasibility of the different wear formation scenarios. Those criteria can be used to rationalize the transition between mild and severe wear regimes and help define the notion of asperity.
				
				\vspace{1em}
				{\footnotesize\noindent\emph{Keywords:} adhesive wear, wear transition, asperity interaction, boundary element method, molecular dynamics}
			}
		\end{center}

		\vspace{1em}

	\end{@twocolumnfalse}
]

\footnotetext{Corresponding author. E-mail address: \href{mailto:son.phamba@epfl.ch}{son.phamba@epfl.ch}}

\section{Introduction}

\begin{table*}[t]
	\centering
	\caption{List of symbols}
	\label{tab:symbols}
	\small
\begin{tabular}{ll}
	\toprule
	\textbf{Symbol}     & \textbf{Description} \\
	\midrule
	$d$                 & Size of micro-contact \\
	$d^*$\hspace*{-0.5em},\hspace{0.2em} $d_\textnormal{a}$, $d_\textnormal{r}$ & Critical, apparent and real contact size \\
	$h$                 & Height of micro-contact in the MD simulations \\
	$a$                 & Distance between point loads \\
	$\lambda$           & Distance between uniform loads \\
	$N$                 & Number of loads \\
	$Q$                 & Point load \\
	$q$                 & Uniform load \\
	$\sigma_\textnormal{j}$          & Shear strength \\
	$\Omega$, $\Omega_\textnormal{d}$ & Two dimensional semi-infinite solid, continuous and discretized \\
	$E$, $G$, $\nu$     & Young's modulus, shear modulus and Poisson's ratio \\
	$\gamma$            & Surface energy \\
	$B$                 & Thickness \\
	$L$, $H$            & Length and height of discretized domain \\
	$n_x$, $n_z$        & Number of discretization points in $x$ and $z$ direction \\
	$\mathcal{M}$       & Spatial discretization factor \\
	$E_\textnormal{el}$, $E_\textnormal{ad}$ & Elastic and adhesive energy \\
	$\mathcal{R}$       & Ratio between elastic and adhesive energy \\
	\hspace{0.1em}$\kappa$            & Ratio between real and apparent contact size \\
	\bottomrule
\end{tabular}
\end{table*}

Wear is a phenomenon happening at nearly every interface between two sliding solids, and yet is far from being fully understood\citewp{rabinowiczFrictionWearMaterials1995}. One of the first and simplest models of wear is Archard's wear model\citewp{archardContactRubbingFlat1953}, giving the wear volume as
\begin{equation}
	\label{eq:archard}
	V = k\frac{F_\textnormal{N} s}{\mathcal{H}}
\end{equation}
where $F_\textnormal{N}$ is the normal load, $s$ the sliding distance, $\mathcal{H}$ the hardness of the softest material and $k$ a fitted wear coefficient. The wear coefficient is usually not constant over time and transitions can be observed between different regimes of wear for different sliding distances or different loads\citewp{zhangTransitionMildSevere1997,rabinowiczLeastWear1984}. Furthermore, the physical origin of the wear coefficient $k$ is not well understood. This motivates the need to obtain a fundamental understanding of wear in order to be able to predict the value of the wear coefficient for a given situation.

A challenge when trying to come up with predictive models for friction and wear\cite{vakisModelingSimulationTribology2018} is that surfaces are typically rough at multiple scales, often with self-affine fractal roughness\citewp{thomNanoscaleRoughnessNatural2017,mandelbrotFractalCharacterFracture1984,majumdarFractalCharacterizationSimulation1990,candelaMinimumScaleGrooving2016}. The contact between two rough surfaces occurs at a small fraction of opposing asperities, and it is critical to assess how large the real contact area is and what pressures are carried at micro-contacts. Rough contact theories estimating these quantities have been derived by considering the interaction of asperities and their statistical distribution on the surfaces\citewp{perssonContactMechanicsRandomly2006,bushElasticContactRough1975,ciavarellaRoughContactsFull2016,greenwoodSurfaceRoughnessContact2001}. Numerical simulations of rough contact between self-affine surfaces generally consider a minimum wavelength in the roughness corresponding to the smallest size of asperities\citewp{hyunFiniteelementAnalysisContact2004,yastrebovInfinitesimalFullContact2015}. This minimum size is physically justified by the roughness vanishing at the atomic scale. Recently, \textcite{frerotMechanisticUnderstandingWear2018} proposed an analytical and numerical mesoscale model relating the distribution of micro-contact sizes to the debris wear volume produced. In yet another recent mesoscale work, the possibility of forming wear particles was discussed in rough contact simulations using an energetic criterion encompassing several nearby micro-contacts\citewp{popovAdhesiveWearParticle2018}.

Such mesoscale wear models need to incorporate a fundamental understanding of when, where and how debris particles are generated at the asperity level. Such a fundamental advance was made for the case of adhesive wear, i.e. when two contacting asperities of materials of similar hardness meet and form an adhesive bond. Molecular dynamics (MD) simulations of two colliding asperities located on opposed surfaces under sliding motion revealed a transition between ductile shearing of asperities to brittle failure and debris formation\citewp{aghababaeiCriticalLengthScale2016,aghababaeiDebrislevelOriginsAdhesive2017,brinkAdhesiveWearMechanisms2019}. This transition occurs at a critical micro-contact size $d^*$ (see Table~\ref{tab:symbols}). Contact junctions of diameter $d$ smaller than this critical size deform plastically upon sliding, while larger junctions form debris. Following the approach used by \textcite{rabinowiczEffectSizeLooseness1958} and \textcite{aghababaeiCriticalLengthScale2016}\snspace, the critical size $d^*$ is found by comparing the elastic energy $E_\textnormal{el}$ stored by shearing the asperities with the adhesive energy $E_\textnormal{ad}$ needed to create the new fractured surfaces that would lead to the formation of a debris particle. If $E_\textnormal{el} \geqslant E_\textnormal{ad}$, there is enough energy to create the new surfaces and therefore form a particle, while if $E_\textnormal{el} < E_\textnormal{ad}$, new surfaces cannot be created and the asperities can only deform plastically. This requirement is similar to Griffith's criterion for the growth of a crack in a material\citewp{andersonFractureMechanicsFundamentals2005}. The expression of the critical size found by \textcite{aghababaeiCriticalLengthScale2016} is
\begin{equation}
	\label{eq:dstar_ramin}
	d^* = \Lambda \frac{4\gamma G}{\sigma_\textnormal{j}^2} \,,
\end{equation}
where $\Lambda$ is a geometrical factor of the order of unity, $\gamma$ is the surface energy of the material, $G$ is the shear modulus and $\sigma_\textnormal{j}$ is the shear strength. $d^*$ depends only on the material properties and not on the geometry, considering the fact that $\Lambda \sim O(1)$. The expression of $d^*$ was later extended\cite{brinkAdhesiveWearMechanisms2019} to take into account a weakened and tilted interface between the joined asperities, resulting in a correction of the term $\sigma_\textnormal{j}$. A distinction is made between having slip or plastic deformations at the junction, but since we are mainly interested in cases of debris formation, we will not consider the two regimes separately in this paper.

The transition between a case where surfaces plastically smoothen and a case where debris particles can be formed is given as a possible explanation for the regimes of low wear and mild wear. For a contact under light normal load, we expect to have a small real contact area between the surfaces, meaning that the junction sizes are small\citewp{yastrebovInfinitesimalFullContact2015}, resulting in no formation of wear particles. For higher loads, some junction sizes become larger than $d^*$\snspace, resulting in the formation of wear particles\citewp{frerotMechanisticUnderstandingWear2018}.

A more recent work\cite{aghababaeiAsperityLevelOriginsTransition2018} has studied how two neighboring pairs of colliding asperities interact, with each pair having a junction size $d \geqslant d^*$\snspace, meaning that each junction can result in the formation of a debris particle under shear. It was shown using 2D MD simulations that when the distance $\lambda$ separating the pairs of asperities is large, each pair of colliding asperities forms its own debris particle of diameter $d$. However, when $\lambda$ becomes smaller than approximately $d$, a single (larger) debris particle combining the two pairs of asperities is formed with an effective diameter of $d_\textnormal{a} = 2d + \lambda$. This change of behavior can be an explanation for the transition from mild wear to severe wear regime at large normal loads, with an increase of the wear coefficient\citewp{rabinowiczLeastWear1984,zhangTransitionMildSevere1997}. The number of micro-contacts increases with load, which promotes the interactions between them and thus the creation of larger debris particles. While no theoretical prediction was given for the critical interaction distance $\lambda$, fracture mechanics can be put forward to rationalize the transition\citewp{aghababaeiAsperityLevelOriginsTransition2018}. The stress intensity factors $K_\text{I}$ of mode I fracture have been analyzed at the corners of each asperity, revealing a mechanism of crack shielding when the pairs of asperities are put closer together\citewp{andersonFractureMechanicsFundamentals2005}, preventing the formation of separate debris particles. Moreover, it is known that multiple neighboring asperities interact elastically over long distances, even without the presence of cracks\citewp{blockPeriodicContactProblems2008,komvopoulosElasticFiniteElement1992a}, with a stress state deviating from the usual Hertzian stress distribution.

\begin{figure}[t]
	\centering
	\includegraphics{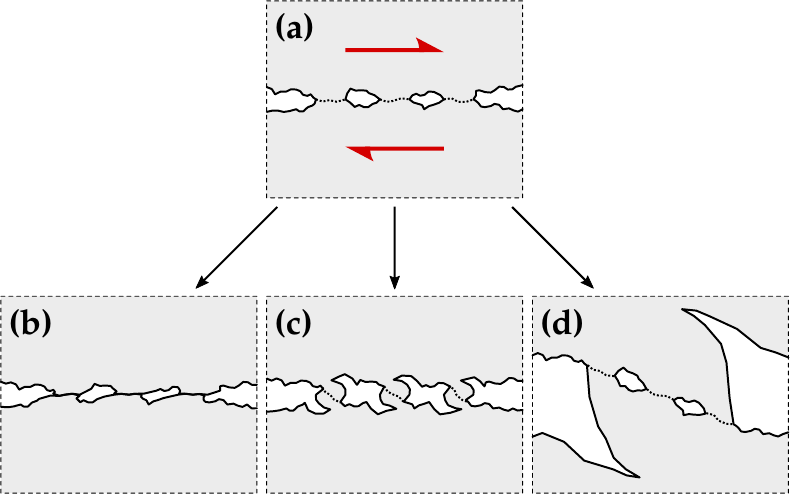}
	\caption[Schematics of sheared micro-contacts and possible outcomes]{Schematics of sheared micro-contacts and possible outcomes. \textbf{(a)}~Initial state. The contact between two solids (top and bottom) viewed at a small scale results in the formation of perfect adhesive junctions (micro-contacts) between the two bodies, shown with dotted lines. We consider equally-spaced micro-contacts of identical sizes. The system is under shear. \textbf{(b)}~Plastic smoothing. \textbf{(c)}~Formation of separate wear particles. \textbf{(d)}~Formation of a combined wear particle.}
	\label{fig:system}
\end{figure}

The objective of this paper is to derive an analytical description of the transitions between different regimes of wear for a given set of colliding asperities. To ease the analytical derivation, we consider flat perfect junctions, or micro-contacts (Fig.~\ref{fig:system}(a)), which are shown to have equivalent properties to colliding asperities regarding the transition between a plastic smoothing regime (Fig.~\ref{fig:system}(b)) and a debris formation regime (Fig.~\ref{fig:system}(c)) at a length scale $d^*$\snspace. An analytical theory for the interaction of multiple micro-contacts is derived in 2D, predicting the transition between a mild wear regime (Fig.~\ref{fig:system}(c)) and a severe wear regime\cite{aghababaeiAsperityLevelOriginsTransition2018} (Fig.~\ref{fig:system}(d)) at the scale of the micro-contacts. The various assumptions are verified using boundary element simulations, and the analytical theory is then validated against MD simulations by simulating perfect adhesive junctions between two solids using model potentials\cite{aghababaeiDebrislevelOriginsAdhesive2017} in quasi-2D setups.

\section{Theoretical model}

We derive an analytical prediction for the outcome of the system shown in Fig.~\ref{fig:system}(a). An energy balance criterion has been effective to predict the transition from plastic shearing to single debris particle creation for both numerical simulations and experimental data\citewp{aghababaeiCriticalLengthScale2016}, therefore the same argument will be used here. The elastic energy is calculated for systems of increasing complexity, starting from a simple point shear loading. We take advantage of the symmetry of the system by only considering the loaded bottom solid, knowing that the top one will be under a symmetric stress state and thus store the same amount of elastic energy. The adhesive energy corresponding to the different outcomes is derived and compared to the stored elastic energy to obtain an energy criterion for the formation of wear particles.

\subsection{Elastic energy}

\paragraph{Point load}

Let us consider a semi-infinite solid in 2D, defined by $\Omega = \{(x, z) \in \mathbb{R}, z \geqslant  0\}$. We call $B$ the thickness of the solid in the $y$ direction. $E$ and $\nu$ are respectively the Young's modulus and the Poisson's ratio of the material.

\begin{figure}[H]
	\centering
	\includegraphics{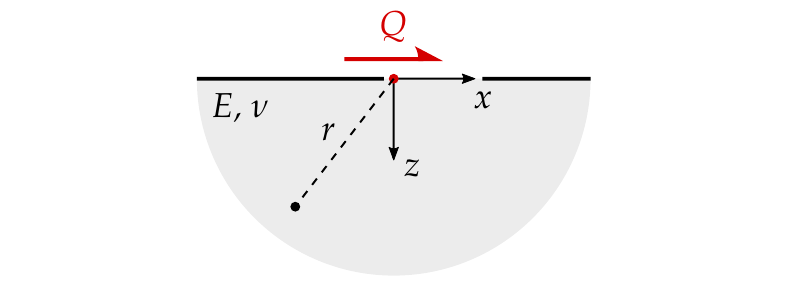}
	\caption[2D semi-infinite solid under a single tangential point load]{2D semi-infinite solid under a single tangential point load. Since the point load $Q$ would be represented by a line load in a 3D equivalent setup, $Q$ has units of force per length.}
	\label{fig:one_point_load}
\end{figure}

A tangentially loaded micro-contact of negligible size can be modeled as a tangential point load of magnitude $Q$ (in units of force per length) applied at the surface of $\Omega$, as shown in Fig.~\ref{fig:one_point_load}.

For a load applied at $x=z=0$, the stresses inside $\Omega$ are\cite{johnsonContactMechanics1985a}
\begin{subequations}\label{eq:stresses_point_load}
\begin{align}
	\sigma_x  &= -\frac{2Q}{\pi} \frac{x^3} {r^4} \,, \\
	\sigma_z  &= -\frac{2Q}{\pi} \frac{xz^2}{r^4} \,, \\
	\tau_{xz} &= -\frac{2Q}{\pi} \frac{x^2z}{r^4} \,,
\end{align}
\end{subequations}
where $r^2 = x^2 + z^2$\snspace[0.4em]. In plane strain conditions, the strains are
\begin{equation}
	\begin{bmatrix}
		\varepsilon_x \\
		\varepsilon_z \\
		\varepsilon_{xz}
	\end{bmatrix}
	= \frac{1 + \nu}{E}
	\begin{bmatrix}
		1 - \nu & -\nu    & 0 \\
		-\nu    & 1 - \nu & 0 \\
		0       & 0       & 1
	\end{bmatrix}
	\begin{bmatrix}
		\sigma_x \\
		\sigma_z \\
		\tau_{xz}
	\end{bmatrix}
	.
\end{equation}
The expression of the elastic energy stored in $\Omega$ is
\begin{align}
	E_\textnormal{el} &= \frac{1}{2} \int_\Omega \bm{\sigma} : \bm{\varepsilon}\, d\Omega \label{eq:Eel_integral} \\
	       &= \frac{1 + \nu}{2E} \int_0^\infty \int_{-\infty}^\infty \left[ (1-\nu)(\sigma_x^2 + \sigma_z^2) \right.\notag\\
	       &\phantom{=\frac{1 + \nu}{2E} \int_0^\infty \int_{-\infty}^\infty [}\left.\vphantom{sigma_x^2} - 2\nu\sigma_x \sigma_z + 2\tau_{xz}^2 \right]B\, dx\, dz \,,\label{eq:Eel_integral_developed}
\end{align}
where $\bm{\sigma} : \bm{\varepsilon}$ is the inner product defined as $\sum_{i, j} \sigma_{ij}\varepsilon_{ij}$. For a load in 2D, the elastic energy is infinite. We can still integrate only along the $x$ direction and keep the infinite term within the integral, which for a tangential point load gives:
\begin{equation}\label{eq:Eel_1Q}
	E_{\textnormal{el},1Q} = \frac{(1 - \nu^2)BQ^2}{\pi E} \int_0^\infty \frac{dz}{z} \,.
\end{equation}
Subsequent expressions of elastic energy can be compared with each other by looking at the factor in front of the integral term. In this case, we notice that the elastic energy is quadratic to the load $Q$.

\paragraph{Infinite integral term}

The fact that the elastic energy stored in a loaded semi-infinite medium is infinite can be explained in several ways. Since the stored elastic energy is equal to the work of the load, it can be calculated by multiplying the magnitude of the load by the displacement of the loaded point in the direction of the load. In 2D, the displacement caused by a load on the surface is $O(\log r)$, meaning that imposing a zero displacement at $r \rightarrow \infty$ as a boundary condition will lead to an infinite displacement under the load\citewp{johnsonContactMechanics1985a}, therefore to an infinite elastic energy. Also, by looking at the stresses in Eq.~\eqref{eq:stresses_point_load}, we see that they are $O(1/r)$, which has a singularity at $r = 0$ and creates the $1/z$ term in the integral of \eqref{eq:Eel_1Q}, decaying too slowly to make the integral finite. The problem of the slow decay is no longer an issue when dealing with systems of finite size. Moreover, the stress singularities disappear in real systems due to plasticity, and as well in simulated systems due to the discretization size. Therefore, \eqref{eq:Eel_1Q} can be rewritten as
\begin{equation}
	E_{\textnormal{el},1Q} = \frac{(1 - \nu^2)BQ^2}{\pi E} \mathcal{M} \label{eq:Eel_1Q_M}
\end{equation}
where $\mathcal{M}$ is a number replacing the infinite integral term, which, again, is finite for a given simulation domain size and discretization size.

Note also that the problem of the infinite elastic energy does not exist in a 3D system (in the absence of stress singularities), because the stresses are $O(1/r^2)$, making the integrals of \eqref{eq:Eel_integral} finite, and the displacements are $O(1/r)$, allowing the application of the boundary condition of zero displacement at $r \rightarrow \infty$.

\paragraph{Two point loads}

\begin{figure}[H]
	\centering
	\includegraphics{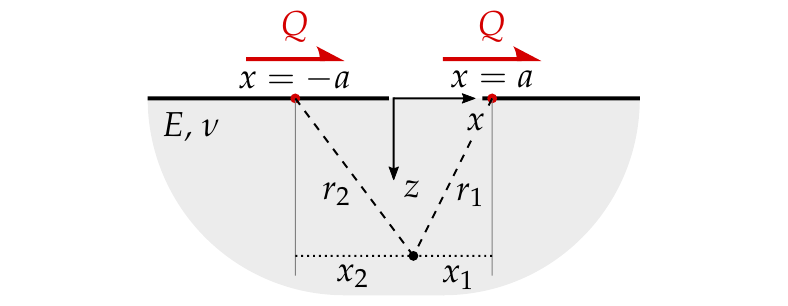}
	\caption[2D semi-infinite solid under two tangential point loads]{2D semi-infinite solid under two tangential point loads. The distance between the two point loads is $2a$.}
	\label{fig:two_point_loads}
\end{figure}

We now consider two tangential point loads, each one of magnitude $Q$, located at $x = a$ and $x = -a$, $z = 0$, as shown in Fig.~\ref{fig:two_point_loads}.

By superposition of (\ref{eq:stresses_point_load}), the stresses are simply
\begin{subequations}
\begin{align}
	\sigma_x  &= -\frac{2Q}{\pi} \left( \frac{x_1^3} {r_1^4} + \frac{x_2^3} {r_2^4} \right)\,, \\
	\sigma_z  &= -\frac{2Q}{\pi} \left( \frac{x_1z^2}{r_1^4} + \frac{x_2z^2}{r_2^4} \right)\,, \\
	\tau_{xz} &= -\frac{2Q}{\pi} \left( \frac{x_1^2z}{r_1^4} + \frac{x_2^2z}{r_2^4} \right)\,,
\end{align}
\end{subequations}
where $x_1 = x - a$, $x_2 = x + a$, $r_1^2 = x_1^2 + z^2$ and $r_2^2 = x_2^2 + z^2$\snspace. The elastic energy is obtained by using these new stresses in \eqref{eq:Eel_integral}, giving:
\begin{align} 
	&E_{\textnormal{el},2Q} = \notag\\
	&\quad\frac{(1 - \nu^2)BQ^2}{\pi E} \left[ 2\int_0^\infty \frac{dz}{z} + 2\int_0^\infty \frac{z}{a^2 + z^2}dz \right]
\end{align}
which is dependent on $a$, the half spacing between the two point loads. If $a$ goes to 0, the two integral terms are the same and the expression of elastic energy is reduced to
\begin{equation}\label{eq:Eel_2Q}
	\left. E_{\textnormal{el},2Q} \right|_{a=0} = 4E_{\textnormal{el},1Q} \,,
\end{equation}
where $E_{\textnormal{el},1Q}$ is the elastic energy of a single point load $Q$ given by \eqref{eq:Eel_1Q_M}. This is consistent with the fact that this situation is equivalent to having a single point load of magnitude $2Q$ (Saint-Venant's principle), the elastic energy being quadratic to the load. If $a$ goes to infinity, the right integral term vanishes, leading to
\begin{equation}
	\lim_{a \to \infty} E_{\textnormal{el},2Q} = 2E_{\textnormal{el},1Q}
\end{equation}
which corresponds to the case where the two point loads are so far apart that they are not interacting, so the total elastic energy is the sum of their individual elastic energy if they were taken separately. Since $a$ always has a finite value compared to the infinite size of the medium, $a$ is dominated by $z$ and the $1/z$ behavior of $z/(a^2 + z^2)$ at infinity, such that $a$ has no effect on the integral if it is not infinite. Therefore, we can assume that we are in the case where $a$ goes to 0, which means that $E_{\textnormal{el},2Q} = 4E_{\textnormal{el},1Q}$. This equality will of course not be exactly matched when performing simulations of finite size, where $a$ will no longer be infinitely small compared to the simulated medium.

\paragraph{$N$ point loads}

We consider $N$ tangential point loads of magnitude $Q$. Following the assumption that the distance $a$ between the loads has a finite value compared to the infinite size of the medium, the setup is equivalent to having a single point load of magnitude $NQ$. The elastic energy being quadratic to the total load, it is also quadratic to the number of point loads:
\begin{equation}\label{eq:Eel_NQ}
	E_{\textnormal{el},NQ} = N^2E_{\textnormal{el},1Q} \,.
\end{equation}
Equation~\eqref{eq:Eel_NQ} is subject to the same limitations as \eqref{eq:Eel_2Q} regarding the simulations of finite size.

\paragraph{Uniform load}

\begin{figure}[H]
	\centering
	\includegraphics{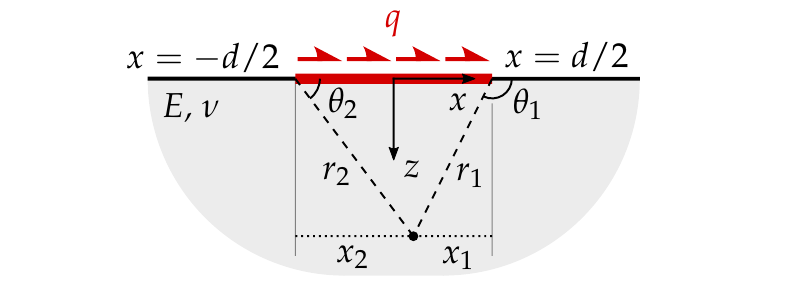}
	\caption[2D semi-infinite solid under a uniform tangential load]{2D semi-infinite solid under a uniform tangential load $q$ of size $d$. Since the load $q$ would be represented by a load on a band in a 3D equivalent setup, $q$ has units of pressure.}
	\label{fig:one_uniform_load}
\end{figure}

A single tangentially loaded micro-contact is better modeled by a uniformly distributed load. We consider a tangential load $q$ (in units of pressure) applied on a region $-d/2 \leqslant x \leqslant d/2$, $z = 0$, as shown in Fig.~\ref{fig:one_uniform_load}, where $d$ is the size of the micro-contact.

Integrating (\ref{eq:stresses_point_load}) on this region, the stresses are
\begin{subequations}
\begin{align}
	\sigma_x  &=  \frac{q}{\pi} \left[ 2\log\left(\frac{r_1}{r_2}\right) - \left( \frac{x_1^2}{r_1^2} - \frac{x_2^2}{r_2^2} \right)\right] \,, \\
	\sigma_z  &= -\frac{q}{\pi} \left( \frac{x_1^2}{r_1^2} - \frac{x_2^2}{r_2^2} \right) \,, \\
	\tau_{xz} &= -\frac{q}{\pi} \left[ (\theta_1 - \theta_2) - \left( \frac{x_1z}{r_1^2} - \frac{x_2z}{r_2^2} \right)\right] \,,
\end{align}
\end{subequations}
where $x_1 = x - d/2$, $x_2 = x + d/2$, $r_1^2 = x_1^2 + z^2$\snspace, $r_2^2 = x_2^2 + z^2$\snspace, $\tan\theta_1 = z_1/x_1$ and $\tan\theta_2 = z_2/x_2$. For the calculation of the elastic energy, the integral (\ref{eq:Eel_integral_developed}) cannot be fully evaluated for this distribution of stresses using analytic functions. However, numerical integration shows that the elastic energy is of the form
\begin{equation}
	E_{\textnormal{el},1q} = \frac{(1 - \nu^2)Bd^2q^2}{\pi E} \int_0^\infty f(z)\,dz
\end{equation}
where $f(z)$ is a function dominated by the term $1/z$ when $z \gg d$. It means that the elastic energy is the same as the one for a point load of equivalent magnitude $Q = dq$:
\begin{equation}
	E_{\textnormal{el},1q} = \left.E_{\textnormal{el},1Q}\right|_{Q = dq} \,,
\end{equation}
in the case where $d$ is small compared to the size of the medium.

\paragraph{$N$ uniform loads}

Using the same assumption as for the $N$ point loads, we can finally derive an expression for the elastic energy of $N$ non-overlapping uniform tangential loads of magnitude $q$ and diameter $d$:
\begin{equation}
	E_{\textnormal{el},Nq} = N^2E_{\textnormal{el},1q} = \frac{(1 - \nu^2)BN^2d^2q^2}{\pi E} \mathcal{M} \,. \label{eq:Eel_Nq}
\end{equation}

This is the expression that will be used to compare with the adhesive energy.

\subsection{Adhesive energy}

\begin{figure}[H]
	\centering
	\includegraphics{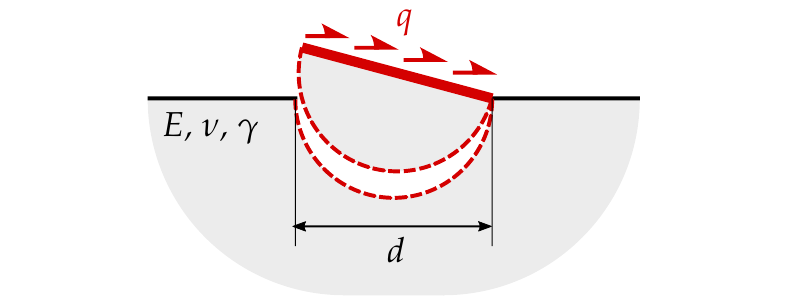}
	\caption[Formation of a debris particle of size $d$ due to the tangential load $q$]{Formation of a debris particle of size $d$ due to the tangential load $q$. Newly created surfaces are shown in red dashed lines.}
	\label{fig:Ead}
\end{figure}

The creation of a debris particle under a micro-contact assuming brittle failure involves the creation of new surfaces. To detach a semi-circular particle of diameter $d$ as shown in Fig.~\ref{fig:Ead}, two surfaces of area $B\pi d/2$ have to be created, which requires an adhesive energy of
\begin{equation}
	E_\textnormal{ad} = \pi \gamma Bd
\end{equation}
where $\gamma$ is the surface energy of the material. The distribution of the stresses forces the crack to initially move into the volume below the surface. We thus assume that the semi-circular path of the crack is close the minimal path that would relieve the stresses in the solid and allow the pivoting of the forming debris particle. Considering another general crack path would only change the adhesive energy up to a small geometrical factor. This assumption is consistent with what is observed in molecular dynamics simulations\citewp{aghababaeiCriticalLengthScale2016,aghababaeiDebrislevelOriginsAdhesive2017}.

\begin{figure}[H]
	\centering
	\includegraphics{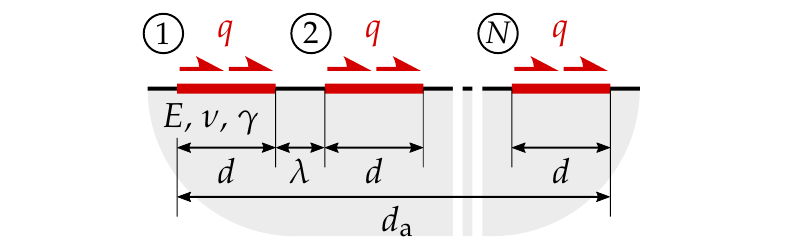}
	\caption[$N$ equally-spaced uniform loads $q$ of size $d$ and spacing $\lambda$]{$N$ equally-spaced uniform loads $q$ of size $d$ and spacing $\lambda$. The total contact size is $d_\textnormal{r} = Nd$ and the apparent contact size is $d_\textnormal{a} = Nd + (N - 1) \lambda$.}
	\label{fig:N_contacts}
\end{figure}

We now consider $N$ equally spaced micro-contacts of size $d$, with $\lambda$ the distance between the edges of two adjacent micro-contacts, as shown in Fig.~\ref{fig:N_contacts}.

Two cases of debris formation can arise. Either $N$ separate debris particles of diameter $d$ are formed, requiring an adhesive energy of
\begin{equation}\label{eq:Eel_ad_sep}
	E_{\textnormal{ad,sep}} = \pi \gamma B d_\textnormal{r} \,,
\end{equation}
where \begin{equation}
	d_\textnormal{r} = Nd
\end{equation}
is the real size of the contact, or in other words the sum of the sizes of all the micro-contacts.

The other possible case of debris formation is to create a single debris particle combining all micro-contacts, having an apparent diameter of
\begin{equation}
	d_\textnormal{a} = Nd + (N - 1) \lambda
\end{equation}
and requiring an adhesive energy of
\begin{equation}\label{eq:Eel_ad_comb}
	E_{\textnormal{ad,comb}}= \pi \gamma B d_\textnormal{a} \,.
\end{equation}
Since we always have $d_\textnormal{a} \geqslant d_\textnormal{r}$, $E_{\textnormal{ad,comb}}$ is always larger than $E_{\textnormal{ad,sep}}$.

\subsection{Energy criterion for debris formation}

\paragraph{Critical micro-contact size}

The formation of debris particles is possible if the stored elastic energy is greater than the adhesive energy required to create the particles\citewp{aghababaeiCriticalLengthScale2016}. In other words, the ratio
\begin{equation}\label{eq:E_ratio}
	\mathcal{R} = \frac{E_\textnormal{el}}{E_\textnormal{ad}}
\end{equation}
has to be greater than one to enable the process of debris formation. For a single micro-contact ($N=1$) of diameter $d$ and uniform load $q$, the energy ratio is equal to
\begin{equation}
	\mathcal{R} = \frac{1 - \nu^2}{\pi^2 \gamma E} dq^2 \mathcal{M} \,.
\end{equation}
If $\mathcal{R} < 1$, no debris particle can be formed under tangential load, so the micro-contact slips or flows plastically. If $\mathcal{R} \geqslant 1$, there is enough energy to create a debris particle. This allows us to define a critical size $d^*$ for a single micro-contact, which is the size above which a debris particle can be created:
\begin{equation}
	d^* = \frac{\pi^2 \gamma E}{(1 - \nu^2) q^2 \mathcal{M}} \,. \label{eq:dstar}
\end{equation}
If we set the tangential load $q$ to be equal to the shear strength of the micro-contact $\sigma_\textnormal{j}$, it becomes clear that $d^*$ is a function of the material parameters, similarly to the critical junction size derived by \citeauthor{aghababaeiCriticalLengthScale2016}, see \eqref{eq:dstar_ramin}. \citeauthor{aghababaeiCriticalLengthScale2016} derived the critical junction size for two spherical colliding asperities, assuming that the elastic energy is stored inside the volume of the asperities and is therefore independent of the system size.  The existence of $d^*$ given by \eqref{eq:dstar} means that, as for colliding asperities, flat junctions also have a ductile-to-brittle transition at a length scale $d^*$\snspace, even if the stored elastic energy is radiated into the bulk.

Note that the expression of $d^*$ \eqref{eq:dstar} is dependent on the number $\mathcal{M}$, which is linked to the spatial discretization and the size of the domain and therefore is not a physical parameter. We recall that $\mathcal{M}$ disappears in a three-dimensional formulation. In reality, $d^*$ is thus a material parameter, to first order independent of geometry. The following arguments will therefore be made in terms of $d^*$ rather than $\mathcal{M}$. The material properties $E$, $\nu$, $\gamma$ and $\sigma_\textnormal{j}$ will all be contained within $d^*$\snspace.

\paragraph{Separated debris particles}

To study the possibility of forming separated debris particles from a distribution of $N$ micro-contacts, we have to calculate from \eqref{eq:Eel_Nq} and \eqref{eq:Eel_ad_sep} the ratio
\begin{equation}
	\mathcal{R}_\textnormal{sep} = \frac{E_{\textnormal{el},Nq}}{E_{\textnormal{ad,sep}}} = \frac{d_\textnormal{r}}{d^*} \,,
\end{equation}
where all material parameters are included in $d^*$\snspace. Note that $\mathcal{R}_\textnormal{sep}$ does not depend on the distance between the micro-contacts. Forming $N$ particles is possible if $\mathcal{R}_\textnormal{sep} \geqslant  1$, or, in terms of $d_\textnormal{r}$, if
\begin{equation}
	d_\textnormal{r} \geqslant d^* \,. \label{eq:d_r_min}
\end{equation}

\paragraph{Combined debris particle}

To study the possibility of forming a combined debris particle from a distribution of $N$ micro-contacts, we calculate from \eqref{eq:Eel_Nq} and \eqref{eq:Eel_ad_comb} the ratio
\begin{equation}
	\mathcal{R}_\textnormal{comb} = \frac{E_{\textnormal{el},Nq}}{E_{\textnormal{ad,comb}}} = \frac{d_\textnormal{r}^2}{d_\textnormal{a}d^*} \,. \label{eq:R_comb}
\end{equation}
Forming a combined particle is possible if $\mathcal{R}_\textnormal{comb} \geqslant 1$, or, in terms of $d_\textnormal{a}$, if
\begin{equation}
	d_\textnormal{a} \leqslant \frac{d_\textnormal{r}^2}{d^*} \,. \label{eq:d_a_max}
\end{equation}

\paragraph{Transition between behaviors}

\begin{figure}[t]
	\centering
	\includegraphics{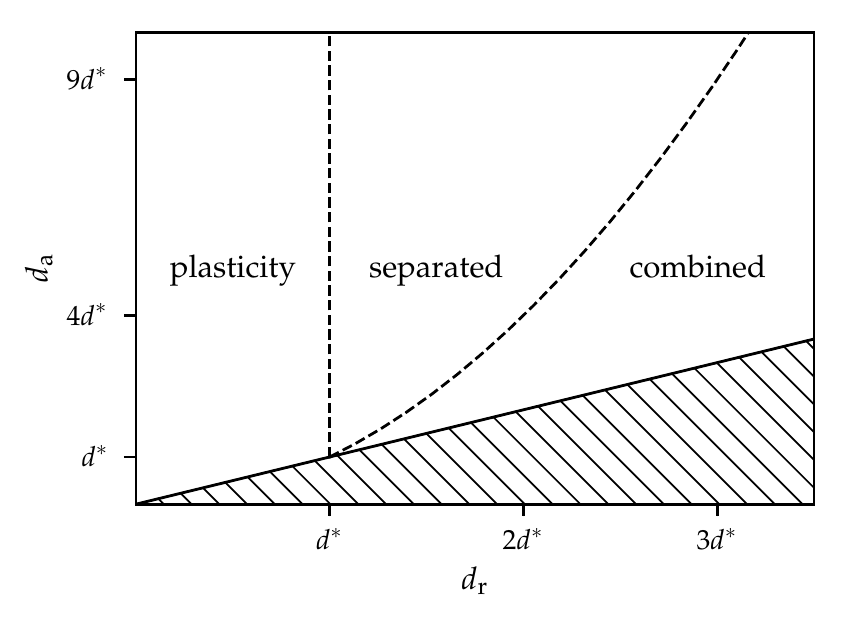}
	\caption{Wear map of the different outcomes for the system of Fig.~\ref{fig:system}(a). The horizontal axis corresponds to the real contact size $d_\textnormal{r}$ and the vertical axis corresponds to the apparent contact size $d_\textnormal{a}$. The hatched area is not accessible because $d_\textnormal{a} \geqslant d_\textnormal{r}$, and its boundary represents systems with a single micro-contact (thus having $d_\textnormal{a} = d_\textnormal{r}$).}
	\label{fig:validation_empty}
\end{figure}

In summary, the three different behaviors described in Fig.~\ref{fig:system} depend only on the values of the real contact size $d_\textnormal{r}$ \eqref{eq:d_r_min} and the apparent contact size $d_\textnormal{a}$ \eqref{eq:d_a_max} and their relation to $d^*$\snspace. Transitions between behaviors are summarized in the wear map shown in Fig.~\ref{fig:validation_empty}, giving regions where each behavior is energetically plausible. The vertical dashed line shows the transition between a plastic behavior and the formation of separated debris particles, as dictated by \eqref{eq:d_r_min}, and the second dashed line shows the transition to the formation of a single combined debris particle, as dictated by \eqref{eq:d_a_max}. The hatched region is inaccessible because $d_\textnormal{a} \geqslant d_\textnormal{r}$.

Our wear map is compatible with the initial definition of a critical length $d^*$ \eqref{eq:dstar_ramin}, found for a single pair of colliding hemispherical asperities (instead of a flat junction of negligible volume, in our case). However, our wear map is not directly applicable to the geometry of colliding asperities when $N \geqslant 2$, because the elastic energy will not only be stored in the volume under the asperities like we assumed, but will mainly be stored in the protruding hemispherical volume of the deforming asperities, which is not taken into account in our model. Therefore, we will expect less elastic interaction in the case of colliding asperities than established for \eqref{eq:Eel_2Q}, since a large part of the energy is contained in the asperities themselves. This results in a lesser likelihood of the formation of combined debris particles. If the asperities are flatter, as expected in many cases of real contacts, the present theory is expected to apply.

\paragraph{Partial contact}

The wear map may be used to define what an asperity is. This definition is far from obvious as, due to the fractal nature of real surfaces, each asperity in contact can be in turn subdivided into smaller contact zones\citewp{archardj.f.ElasticDeformationLaws1957}. What appears as a fully compact contact junction at a given scale, becomes fragmented into smaller contact patches at a smaller scale. Yet, interactions between these divided contact spots may be homogenized into a single apparent contact junctions if elastic interactions prevail, which is precisely what Fig.~\ref{fig:validation_empty} can help assess. For a contact junction of apparent diameter $d_\textnormal{a}$, we can define the fraction of the real contact size to the apparent contact size as
\begin{equation}
	\kappa = \frac{d_\textnormal{r}}{d_\textnormal{a}} \,,
\end{equation}
satisfying $\kappa \leqslant 1$. We establish a criterion to determine if this weakened contact can form a single debris particle of size $d_\textnormal{a}$ by rewriting the condition for combined debris particle formation \eqref{eq:d_a_max} using $\kappa$, leading to the condition
\begin{equation}
	\kappa \geqslant \sqrt{\frac{d^*}{d_\textnormal{a}}} \,,
\end{equation}
which is the minimum fraction of contact size necessary to be able to detach a single debris particle. This minimum fraction is only reachable if $\sqrt{d^*/d_\textnormal{a}} \leqslant 1$, or if
\begin{equation}
	d_\textnormal{a} \geqslant  d^* \,.
\end{equation}
It implies that a large contact of apparent contact size $d_\textnormal{a} \geqslant d^*$ can be broken down into smaller micro-contacts of total real contact size $d_\textnormal{r} = \kappa d_\textnormal{a}$ with $\kappa \geqslant \sqrt{d^*/d_\textnormal{a}}$ and still form a single debris particle from an energetic point of view. It also shows that only the total contact size $d_\textnormal{r}$ matters to determine the formation of debris particles, and not the individual sizes of each micro-contact (assuming full elastic interaction between the micro-contacts).

\paragraph{Limits of the energy criterion}

One may have noticed that we always have $\mathcal{R}_\textnormal{sep} \geqslant  \mathcal{R}_\textnormal{comb}$, meaning that forming small individual debris particles is always more energetically favorable than forming a single, combined one, even if $\mathcal{R}_\textnormal{comb} \geqslant  1$. In this case, the energy criterion will only suggest that both cases are energetically possible, but will not indicate which one will happen. The outcome can be predicted by looking at the locations of stress concentrations in the material, which indicate the places where cracks can nucleate and thus where debris formation can occur. It implies that in the wear map (Fig.~\ref{fig:validation_empty}), a separated debris particle formation can energetically happen in the ``combined'' region.

\section{Validation using simulations}

\subsection{Boundary element method}

\begin{figure}[t] 
	\centering
	\includegraphics{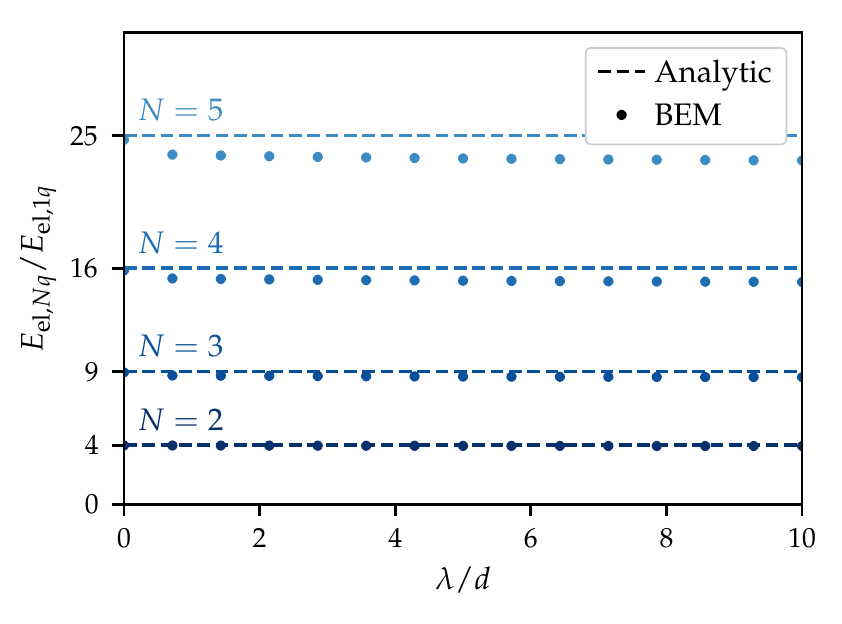}
	\caption[BEM simulations of the elastic energy $E_{\textnormal{el},Nq}$ of $N$ equally-spaced uniform loads $q$ compared to analytical prediction]{BEM simulations of the elastic energy $E_{\textnormal{el},Nq}$ of $N$ equally-spaced uniform loads $q$ compared to analytical prediction. The values were computed with $L = 128$, $H = 2048$, $B = 1$, $n_x = n_z = 2048$, $d = 1$, $E = 1$, $\nu = 0.3$, $q = 0.001$. The energies are normalized by $E_{\textnormal{el},1q}$, computed with a single uniform load, to check the validity of \eqref{eq:Eel_NQ}. The agreement is better when the apparent contact size $d_a$ is small compared to $L$ and $H$.}
	\label{fig:BEM_validation}
\end{figure}

Before validating the wear map (Fig.~\ref{fig:validation_empty}), the boundary element method (BEM) is used to verify the effects of a finite size system on the computation of the elastic energy $E_{\textnormal{el},Nq}$ \eqref{eq:Eel_Nq}, which was derived assuming an infinite medium. Instead of a semi-infinite plane $\Omega$, a region $\Omega_\textnormal{d} = \{|x| \leqslant L/2,\, 0 \leqslant z \leqslant H\}$ of $n_x \times n_z$ points is considered. The BEM simulation is periodic in the $x$ direction. Desired loads are applied at the nodes located at $z = 0$, and the elastic energy is computed by discretizing the integral of (\ref{eq:Eel_integral}) as
\begin{equation}
	E_\textnormal{el} = \frac{1}{2} \sum_{i=0}^{n_x-1} \sum_{j=0}^{n_z-1} \bm{\sigma}(x_i, z_j) : \bm{\varepsilon}(x_i, z_j)\, \frac{L}{n_x} \frac{H}{n_z} \,.
\end{equation}
The stress and strain tensors $\bm{\sigma}(x_i, z_j)$ and $\bm{\varepsilon}(x_i, z_j)$ are evaluated using a code based on the application of Green's functions defined for a tangential point load applied at the surface of a semi-infinite medium\citewp{polonskyNumericalMethodSolving1999,reyNormalAdhesiveContact2017}. The comparison of the results from the BEM with the analytical expression of $E_{\textnormal{el},Nq}$ is shown in Fig.~\ref{fig:BEM_validation}. $H$ is chosen to be large compared to the maximum $d_\textnormal{a}$ (with $H/d_\textnormal{a} > 45$) to match the assumption of a large medium made for the analytical prediction. Having periodicity in the $x$ direction and computing the elastic energy only between $-L/2$ and $L/2$ does not seem to affect the match between the analytical and numerical $E_{\textnormal{el},Nq}$, even if $d_\textnormal{a}$ is not significantly small compared to $L$ (we always have $L/d_\textnormal{a} > 2.8$). No improvement was found by increasing the value of $L$.

Overall, there is a good match between the analytical predictions and the numerical values. The discrepancy between the computed and theoretical values is the smallest for lower values of $\lambda$ and $N$ (low $d_\textnormal{a}$). The decrease of the elastic energy recorded when $\lambda$ increases can be seen as a transition between a complete interaction of the $N$ loads at $\lambda = 0$, and an absence of interaction for $\lambda \rightarrow \infty$. At complete interaction, $E_{\textnormal{el},Nq}$ is verified to be quadratic to $N$, while for no interaction, it should decrease to become proportional to $N$, which is not seen in Fig.~\ref{fig:BEM_validation} because of the small value of $\lambda$.

The theoretical elastic energy slightly overestimates the value computed inside a finite domain, but it still provides a good approximation. Therefore, the criteria for debris formation we derived using the analytical expression of elastic energy can reasonably be applied for systems of finite size.

\subsection{Molecular dynamics}

\begin{figure}
	\centering
	\includegraphics{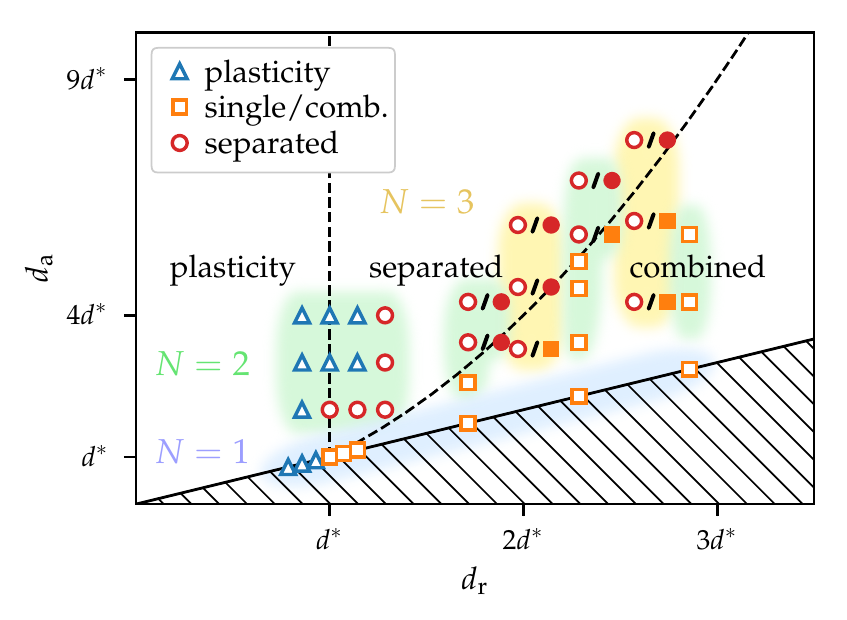}
	\caption[Distribution of the MD simulations' outcomes on the wear map (Fig.~\ref{fig:validation_empty})]{Distribution of the MD simulations' outcomes on the wear map (Fig.~\ref{fig:validation_empty}). Each symbol corresponds to one MD simulation. Empty and color-filled symbols are simulations with sharp corners and rounded corners respectively, sometimes leading to a different outcome. Rounding the corners has the effect of slightly increasing $d_\textnormal{r}$, shifting the symbol to the right in the diagram. The highlighted areas show the different values of $N$ used for the simulations.}
	\label{fig:validation}
\end{figure}

We use MD simulations of perfect junctions to check the validity of the predictions for debris formation, \eqref{eq:d_r_min} and (\ref{eq:d_a_max}). A model potential (P4 in \textcite{aghababaeiDebrislevelOriginsAdhesive2017}) is used to simulate a material brittle enough to have a critical length $d^*$ observable at the scale of the simulations while maintaining a reasonable size. We have $d^* \simeq 35 r_0$ in all our simulations, where $r_0$ is the interatomic distance at absolute zero temperature. The size of all the simulations is kept constant at $L = 900r_0$ in the $x$ direction and $H = 1200r_0$ in the $z$ direction. The atoms are arranged in a face-centered-cubic lattice with a thickness of three close-packed layers of atoms in the $y$ direction to stay in a quasi-2D plane strain representation. The [111] lattice direction is aligned with the $y$ axis and the [110] direction is in the $(x, z)$ plane at an angle of 15$^\circ$ with the $x$ axis, so that the junctions are not aligned with weak crystal planes. Periodic boundary conditions are used in the $x$ and $y$ directions and the possible lattice mismatch at the boundary is resolved using a step of energy minimization. The temperature of the system is kept constant using Langevin thermostats at the non-periodic boundaries of the simulation. Instead of applying a shear force, a constant velocity is imposed on the boundary of the top solid, and a small constant normal load is applied to prevent the system from drifting apart. The maximum resultant shear force is limited by the shear strength $\sigma_\textnormal{j}$ of the material. All the MD simulations are performed with LAMMPS\cite{plimptonFastParallelAlgorithms1995} and visualized with OVITO\citewp{stukowskiVisualizationAnalysisAtomistic2009}.

\begin{figure*}
	\centering
	\includegraphics[center]{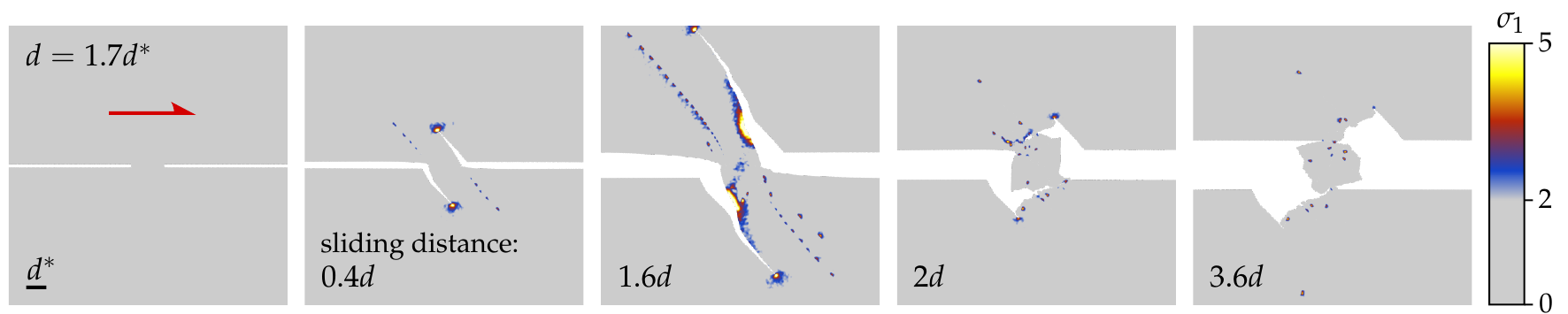}
	\caption[Formation of a debris particle from a single micro-contact of size $d = 1.7d^*$]{Formation of a debris particle from a single micro-contact of size $d = 1.7d^*$\snspace. The red arrow indicates the shear direction. The bottom boundary is fixed. The colors show the first principal stress $\sigma_1$ (if positive: maximum tensile stress) in reduced dimensionless Lennard-Jones units, with brighter regions corresponding to higher values. Regions of stress concentration are visible near the tips of the growing cracks and at the locations of crack nucleation. The smaller bright spots are mismatches in the crystallographic structure propagating in the material, which is the nanoscale manifestation of shear plasticity. Notice that the cracks extend way past the necessary length before closing at the formation of the debris particle, which is due to an excess of stored elastic energy ($\mathcal{R} > 1$).}
	\label{fig:md_n_1}
\end{figure*}

The critical length $d^*$ is first found by simulating single junctions of increasing sizes $d$. The perfect junctions are modeled by rectangles of width $d$ and fixed height $h = 6r_0$. The observation of a transition between plastic behavior and debris particle formation validates the existence of the critical size $d^*$ \eqref{eq:dstar}, which is defined as the value of $d$ at which this transition occurs. The points in Fig.~\ref{fig:validation} show that these simulations are located on the line $d_\textnormal{a} = d_\textnormal{r}$ and are highlighted in light blue. An example of debris particle formation is shown in Fig.~\ref{fig:md_n_1}. Two types of damaging processes are witnessed during the formation of the debris particle: cracks are opened under tensile stress, and dislocations move across the material, which is a nanoscopic manifestation of plastic shear deformations. When varying $d$, the length reached by the opening cracks becomes larger when $d$ increases relative to $d^*$\snspace. This phenomenon is enabled by $\mathcal{R}$ increasing with $d$ (and being greater than 1 because $d > d^*$), meaning that more elastic energy is stored than what is needed to create the minimal crack path. This excess energy can be used in the formation of larger cracks.

\begin{figure}
	\centering
	\includegraphics{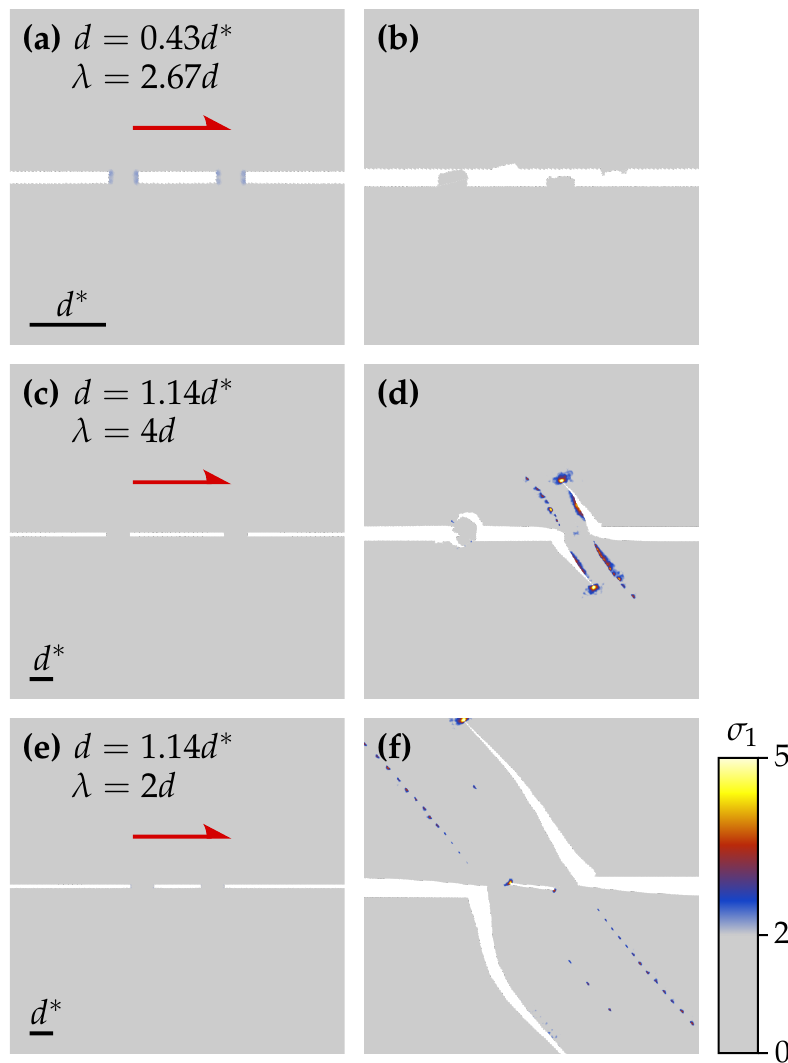}
	\caption[Three different outcomes from two sheared micro-contacts]{Three different outcomes from two sheared micro-contacts. \textbf{(a)}, \textbf{(c)} and \textbf{(e)} show the initial states of \textbf{(b)}, \textbf{(d)} and \textbf{(f)} respectively. \textbf{(b)} Case of slip. \textbf{(d)} Case of separated debris particle formation. The left one is detached and the right one is under formation. \textbf{(f)} Case of combined debris particle formation. The length of the cracks is a characteristic of this behavior.}
	\label{fig:md_n_2}
\end{figure}

Several simulations were performed with $N = 2$ (highlighted in green in Fig.~\ref{fig:validation}) and $N = 3$ (highlighted in orange) with different values of $d$ and $\lambda$. Examples of the three possible outcomes are shown in Fig.~\ref{fig:md_n_2} for $N = 2$. We recall that it is possible to observe the formation of separated debris particles even when the formation of a combined debris particle is energetically possible ($\mathcal{R}_\textnormal{sep} \geqslant \mathcal{R}_\textnormal{comb}$). Therefore, when an MD simulation leads to the formation of separated debris particles, it is relaunched with the same geometric parameters but adding rounded corners where the cracks forming separated debris particles can initiate, in order to prevent their formation, leading to a different outcome if energetically possible (shown by color-filled symbols in Fig.~\ref{fig:validation}). This shows that the outcome is also controlled by the presence or absence of stress concentrations.

Some simulations reach an asymmetric end state even if the initial state seems symmetrical, like shown in Fig.~\ref{fig:md_n_2}(d). This is due to the system being very sensitive to stress concentrators (sharp corners) where even the presence or absence of a single atom can make the system's evolution asymmetrical. This is a side effect of using a brittle interatomic potential for the simulations. The fact of imposing a shearing velocity only on the upper half of the system can also contribute to the asymmetry. This is still consistent with our model since particles form as predicted.

The prediction of the transition between a separated and a combined debris particle formation behavior \eqref{eq:d_a_max} is well matched by the $N=2$ and $N=3$ simulations, as no combined particles are ever formed outside of the ``combined'' region of the wear map. Some simulations with sharp corners lead to a behavior consuming less energy, but their rounded corners counterparts are in agreement with the wear map.

The prediction of the transition between a plastic behavior and the formation of separated debris particles \eqref{eq:d_r_min} is however not perfectly matched by the $N=2$ simulations when $d_\textnormal{a}$ increases, as seen in Fig.~\ref{fig:validation} near the $d_\textnormal{r} = d^*$ dashed line. It means that when the distance between the micro-contacts $\lambda$ gets bigger, there is not enough energy to create the separated debris particles. The decrease of the elastic energy for increasing values of $\lambda$ was identified with the BEM simulations (Fig.~\ref{fig:BEM_validation}) and is also present in the MD simulations, because they both take place in a finite discretized medium. In the wear map (Fig.~\ref{fig:validation}), taking the decrease of elastic energy into account for increasing values of $\lambda$, thus increasing values of $d_\textnormal{a}$, would be represented by shifting the dashed lines (showing the transitions between behaviors) to the right for the higher values of $d_\textnormal{a}$. The disagreement of the MD simulations with our theory near $d_\textnormal{r} = d^*$ might also be an effect of having the sizes of the individual micro-contacts $d \simeq 0.5 d^*$ being close to their height $h = 0.17d^*$\snspace, enabling unwanted geometrical effects like the concentration of elastic energy in the now non-negligible volume of the micro-contacts, resulting in even less elastic interaction between the micro-contacts.

In general, the criteria \eqref{eq:d_r_min} and \eqref{eq:d_a_max} work well to predict the transitions of behavior, and the transition between the separated and combined debris formation is especially well matched by the MD simulations.

\section{Conclusion}

We derived and validated analytical criteria for the formation of separated or combined debris particles in an adhesive wear regime at the microscale, leading to a wear map of the different behaviors. The outcome is dictated by the sum of the sizes of the micro-contacts, i.e. the real contact area, and by the total length covered by all of them, i.e. the apparent contact area, in comparison to the critical length scale $d^*$ of the material, at which a ductile-to-brittle transition occurs. The different microscopic behaviors of debris particle formation give a physical interpretation for the different regimes of macroscopic unlubricated adhesive wear, and the emergence of a regime of severe wear can be physically explained by the energetic feasibility of forming combined debris particles under multiple micro-contacts. Future work will generalize these findings in a 3D setting.

\atColsBreak{\vskip10pt}
\printbibliography

\end{document}